\documentclass[conference]{IEEEtran}
\usepackage[letterpaper,margin=1in]{geometry}
\IEEEoverridecommandlockouts

\usepackage{cite}
\usepackage{amsmath,amssymb,amsfonts}
\usepackage{algorithmic}
\usepackage{algorithm}
\usepackage{graphicx}
\usepackage{textcomp}
\usepackage{xcolor}
\usepackage{booktabs}
\usepackage{multirow}
\usepackage{siunitx}
\usepackage{tabularx}
\usepackage{adjustbox}
\usepackage{threeparttable}
\usepackage{makecell}
\usepackage{array}
\usepackage{enumitem}
\usepackage{hyperref}
\usepackage{microtype}

\newcommand{\framework}{\textsc{NOPE-HYPE}}
\newcommand{\simname}{\textsc{SIM}}
\newcommand{\distcorr}{d_{\mathrm{corr}}}
\newcommand{\RMS}{\mathrm{RMS}}

\def\BibTeX{{\rm B\kern-.05em{\sc i\kern-.025em b}\kern-.08em
    T\kern-.1667em\lower.7ex\hbox{E}\kern-.125emX}}

\begin{document}

\title{NOPE-HYPE: A Structured Simulation Workflow for Robust Speech-to-Text Across Diverse Acoustic Environments}

\author{
\IEEEauthorblockN{Niramay M. Patel}
\IEEEauthorblockA{IISER Bhopal, India \\
niramay23@iiserb.ac.in}
\and
\IEEEauthorblockN{Bibek Behera}
\IEEEauthorblockA{IIT Bombay, India \\
bibek.iitkgp@gmail.com}
\and
\IEEEauthorblockN{Raksha Sharma}
\IEEEauthorblockA{IIT Roorkee, India \\
raksha.sharma@cs.iitr.ac.in}
}

\maketitle

\begin{abstract}
Robust speech-to-text translation systems should perform reliably across diverse acoustic conditions, yet practical pipelines lack controllable tools for systematic environment exploration. Large speech models remain sensitive to unseen acoustic conditions, as training data rarely cover the full range of real environments. We present \framework, a structured training workflow that combines a controllable environment simulator, coverage-optimal environment reduction on Power Spectral Density (PSD) templates, and a small, interpretable hyperparameter search over simulator knobs. We show that simulator-generated noise achieves performance comparable to
balanced real-noise training across Whisper and SeamlessM4T models, provide principled environment prototype sets, and identify practical default simulator configurations from a structured 27-run hyperparameter sweep.
\end{abstract}


\section{Introduction}
\label{sec:intro}

Speech-to-text (S2TT) translation systems often operate outside curated studio audio settings. The audio of the deployment often includes transportation noise, indoor hums, transient events, and speech-like interference from nearby people. Even strong speech foundation models can degrade under such shifts because training data and augmentation choices rarely cover the full space of real acoustic variability in a controlled way. As shown in Table~\ref{tab:main-robustness}, models trained only on clean speech perform worse than those trained with noise injection.

A common mitigation is \emph{noise augmentation}. In practice, this collapses into a single axis, such as Gaussian noise, a single colored-noise type, or a narrow SNR range \cite{ko15_interspeech,hannun2014deepspeechscalingendtoend}. These choices are convenient, but they do not isolate factors that matter in real environments --- the spectral shape differs by scene, energy changes over time, transient activity varies, and human activity adds interference that overlaps with speech. Learned audio generators can increase realism, but their latent factors are often hard to interpret and sweep independently, limiting their usefulness for a targeted robustness search \cite{oord2016wavenet,kumar2019melgan,liu2023audioldm}.

In this paper, we shift the focus from adding arbitrary noise to systematically modeling environment-specific noise to build an S2TT system that performs well under both clean and noisy conditions. We propose \framework, a practical workflow designed to improve systematic robustness through \emph{interpretable control} and \emph{structured search}. We test our framework on two language pairs: English to Hindi and English to German. The workflow has three parts:

\begin{itemize}[leftmargin=*, itemsep=0pt, parsep=0pt, topsep=2pt]
  \item \emph{SIMULATOR (SIM):} A parameterized simulator for environmental noise and noise-in-speech mixtures. Rather than aiming for exactly realistic noise, it enables structured modeling of acoustic factors relevant to speech-to-text performance.
  \item \emph{Environment Reduction:} When many environmental categories exist, uniform sampling becomes computationally expensive because each environment requires a full hyperparameter sweep of the simulator. We summarize each environment using a PSD template and select a small subset of prototypes that covers the full set under explicit distances.
  \item \emph{Hyperparameter Search:} We perform a sequential and interpretable search over simulator parameters that yield a robust default configuration and clarify which acoustic factors strongly influence model performance.
\end{itemize}

The paper is organized as follows. Section~\ref{sec:related} reviews robust speech recognition and augmentation. Section~\ref{sec:overview} summarizes the \framework\ workflow, including the controllable noise simulator (\simname) and environment reduction. Section~\ref{sec:exp} describes experiments evaluating Whisper and SeamlessM4T, including ablations and a 27-run hyperparameter sweep. Sections~\ref{sec:discussion} and~\ref{sec:conclusion} discuss the results and summarize the main findings.

\section{Related Work}
\label{sec:related}

Robust automatic speech recognition (ASR) has often used acoustic augmentation to improve performance in noisy conditions, including additive noise \cite{ko15_interspeech,hannun2014deepspeechscalingendtoend}, reverberation \cite{7953152}, speed and pitch perturbation \cite{ko15_interspeech}, and time-frequency masking \cite{Park_2019}. SpecAugment \cite{Park_2019} established structured masking as a strong baseline, with later extensions to mixed-sample and hidden-space augmentation \cite{kim2021specmixmixedsample,wang2021specaugmenthiddenspacedata}. However, these methods typically rely on coarse heuristics and do not allow systematic exploration of environment structure.

Synthetic data pipelines generate diverse acoustic conditions via simulated noise and reverberation, using either simple RIR/noise addition \cite{ko15_interspeech,7953152} or generative models for realistic noise synthesis \cite{fazel2021synthasrunlockingsyntheticdata,casanova2023asrdataaugmentationlowresource}. While effective for robustness, they often trade simplicity for controllability, making structured ablation difficult \cite{liu2025pretrainingrobustasrfoundation}.

Large-scale S2TT systems such as Whisper \cite{radford2022robustspeechrecognitionlargescale} and SeamlessM4T \cite{communication2023seamlessmultilingualexpressivestreaming} achieve strong baseline robustness, but remain sensitive to training distributions and unseen noise. Evaluations typically focus on fixed benchmarks rather than systematic augmentation design or environment reduction, leaving limited understanding of how environment structure impacts robustness.

\section{Overview of \framework}
\label{sec:overview}

\framework\ is a structured workflow for improving and analyzing acoustic robustness in speech-to-text systems, where robustness refers to maintaining strong performance under changing noise conditions and environment types. Instead of treating noise augmentation as a fixed pre-processing step, it decomposes robustness into three controllable stages.

We use real environmental recordings from the DEMAND dataset \cite{thiemann2013demand} with the environment categories listed in supplementary material. First, each environment is summarized by a Power Spectral Density (PSD) template on a shared frequency grid (SFG). A SFG is used so that all environments are represented on the same frequency bins, allowing their PSD curves to be compared directly. Second, when many environment categories exist, we reduce the set by selecting $k$ prototype environments using a coverage objective over PSD templates. Third, we train with \simname\ using fitted targets from selected environments, generating controlled environment noise mixed with clean speech at sampled SNR levels, and run structured sweeps over simulator parameters to select robustness-oriented configurations. As summarized in Algorithm~\ref{alg:pipeline}, this workflow enables direct comparison across environments under a unified training protocol.

\begin{algorithm}[t]
\caption{Robustness Workflow}
\label{alg:pipeline}
\begin{algorithmic}[1]
\STATE Compute PSD template $x_e$ for each environment $e$
\STATE Optionally select few noise environments by coverage optimization on $\{x_e\}$
\FOR{each environment $e$ in $P$ (or all environments)}
  \STATE Fit simulator targets: PSD, envelope stats, transient rate, optional tonal peaks
\ENDFOR
\FOR{each simulator configuration $\theta$ (default or sweep)}
  \STATE Generate noise with simulator using $\theta$
  \STATE Mix with clean speech at sampled SNR (optional RIR)
  \STATE Train S2TT model with fixed compute budget
  \STATE Evaluate on clean, real-noise, and sim-noise test sets
\ENDFOR
\STATE Select $\theta$ by robustness objective and report ablations
\end{algorithmic}
\end{algorithm}

\section{SIM: A Controllable Environment Audio Simulator}
\label{sec:sim}

The simulator supports robust training and structured parameter search while providing direct control over key acoustic factors. It produces deterministic outputs under a fixed seed, enabling reliable ablations and controlled comparisons. It targets realistic environmental noise by matching stable statistical properties of real audio rather than trying to sound indistinguishable from real recordings.

\subsection{Interface and Modes}
\label{subsec:sim-interface}

\simname\ provides three commands: \texttt{gen} (ambience-only), \texttt{mix} (speech plus ambience at target SNR), and \texttt{eval} (noise similarity statistics). Synthesis uses $f_s = 16\,\mathrm{kHz}$ by default. It supports two modes:

\begin{itemize}[leftmargin=*, itemsep=0pt, parsep=0pt, topsep=2pt]
  \item \emph{Preset mode:} A fully procedural synthesis pipeline built from layered primitives. It generates a colored-noise bed (white, pink, or brown), applies multi-band spectral shaping with per-band gain control, and introduces slow stochastic modulation to create nonstationarity. Event primitives such as band-limited transients or ramps can also be added. This mode requires no real recordings and provides full parameter control, making it suitable for bootstrap experiments and factorized sweeps.
  \item \emph{Generated mode:} Anchors synthesis to real recordings through targets estimated from real ambience, including a mean PSD curve, envelope log-RMS statistics with a slow-dynamics rate hint, transient-rate estimates from onset peaks, and optional tonal peaks. This mode aims to match the stable spectral and temporal properties of real environments while retaining explicit control over each component.
\end{itemize}

\begin{figure}[t]
    \centering
    \includegraphics[width=0.95\columnwidth]{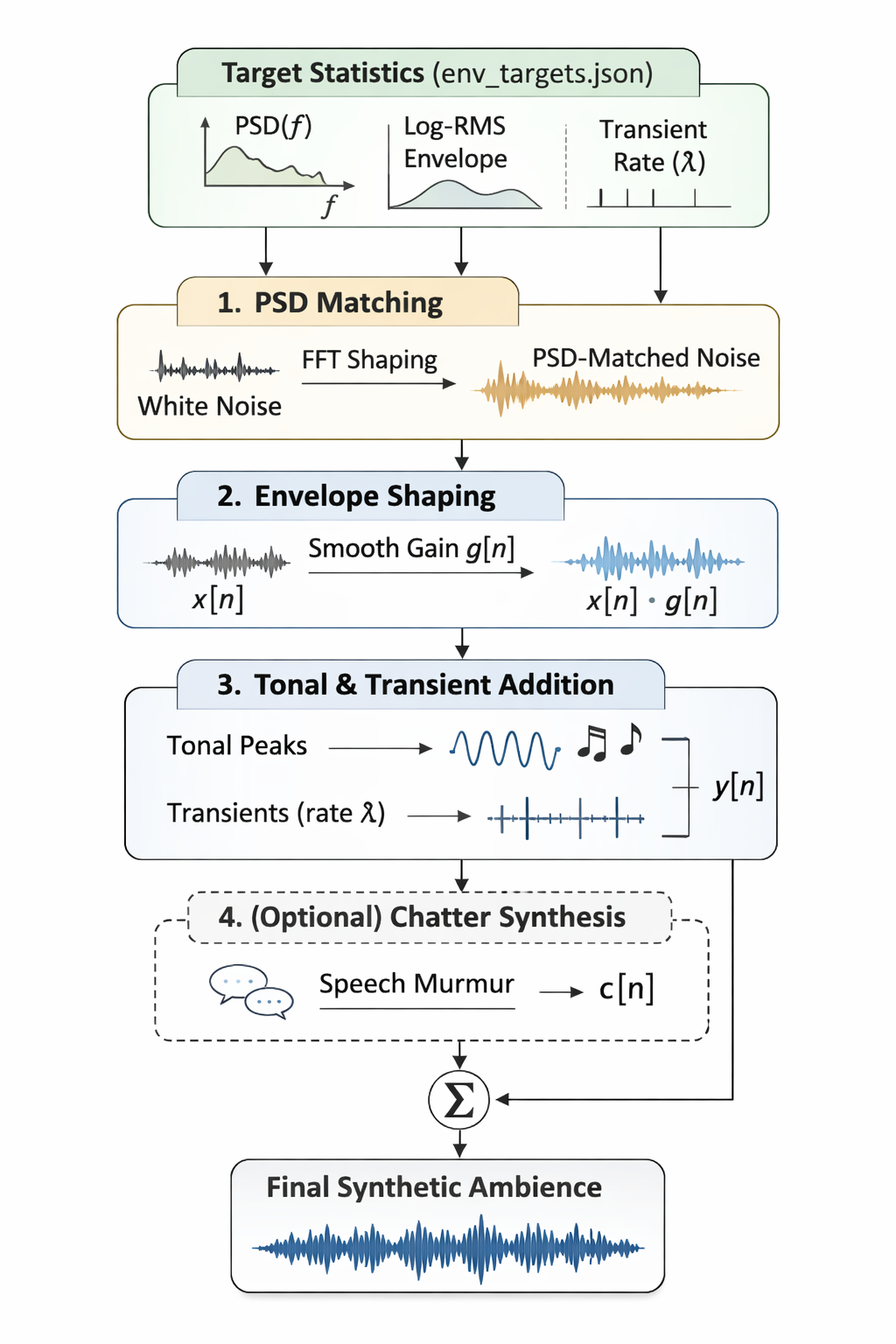}
    \caption{SIM model pipeline.}
    \label{fig:sim_model}
\end{figure}

\subsection{Generated Mode Core Building Blocks}
\label{subsec:sim-blocks}

The generated mode synthesizes noise from real-environment recordings by matching PSD structure, envelope dynamics, transient activity, chatter, and RIR. Fig.~\ref{fig:sim_model} summarizes the pipeline.

\textbf{PSD estimation and PSD-matched bed.}
We estimate a target PSD using Welch averaging \cite{welch1967fftpsd} and synthesize a stationary background bed by matching this target spectrum in the Fourier domain, using target magnitudes with randomized phases and a small number of calibration passes to reduce residual PSD mismatch. This design is motivated by the fact that, for stationary or slowly varying background noise, the PSD captures the dominant spectral structure of the environment, while phase randomization provides a simple way to generate noise in the PSD. During calibration, the PSD correction ratios are clipped to bounded ranges to maintain numerical stability and reproducibility.

\textbf{Envelope shaping.}
We apply a smooth multiplicative envelope $a(t)$ to the stationary PSD-matched bed to model nonstationarity. We generate a low-frequency control signal
\[
u(t) = \mathrm{LPF}_{r_e}(\epsilon(t)),
\]
where $\epsilon(t)$ is white noise and $\mathrm{LPF}_{r_e}(\cdot)$ denotes low-pass filtering with rate parameter $r_e$. The resulting signal is then exponentiated to produce a positive envelope,
\[
a(t) = \exp(\sigma_e u(t)),
\]
where $r_e$ controls how quickly the envelope varies and $\sigma_e$ controls the magnitude of energy fluctuations. This stage allows the simulator to improve beyond a purely stationary background bed and capture the slow amplitude variation observed in real environments.

\textbf{Transient injection.}
Real environments contain short impulsive events such as clicks or impacts. We model these by injecting band-limited transient bursts at a target rate $\lambda$ events per second, shaped by a smooth envelope and filtered within an environment-specific frequency band.

\textbf{Chatter injection.}
We model human background activity by mixing short speech segments sampled from a separate speech pool, band-limited and scheduled in burst-like intervals to mimic background murmur or conversation.

\textbf{RIR.}
For indoor environments, we optionally apply a room impulse response during mixing to simulate reverberant propagation. This introduces reflections characteristic of enclosed spaces and provides explicit control over reverberation.

\subsection{Mixing}
\label{subsec:mixing}

Given clean speech $s(t)$ and simulated ambience $x_e(t)$, we form a noisy utterance at a target SNR:
\[
y(t) = s(t) + g\,x_e(t),
\]
where the scaling factor is
\[
g = \frac{\RMS(s)}{\RMS(x_e)} \, 10^{-\mathrm{SNR}/20}.
\]
so that the resulting mixture matches the target SNR. The derivation of this expression is provided in supplementary material. After mixing, we apply mild soft clipping and peak normalization to keep the waveform within a stable numeric range and avoid occasional amplitude spikes. 

\section{Reducing Environment Sets}
\label{sec:proto}

When many environments are available, sweeping simulator hyperparameters across all environments becomes computationally expensive and often redundant due to similar spectral structure. We reduce the environment set by treating it as a coverage problem in PSD-template space and selecting representative prototypes.

\textbf{PSD templates and distances.}
Each environment $e_i$ is represented by a centered PSD template $x_i \in \mathbb{R}^F$ on a SFG. We compute distances between PSD templates using two complementary measures:
\begin{itemize}[leftmargin=*, itemsep=2pt]
  \item \emph{Correlation distance:} This distance captures whether two environments have similar spectral patterns. 
  \[
   \distcorr(x,y) = 1 - \mathrm{corr}(x,y).
  \]
  \item \emph{RMSE:}  This distance measures how close two PSD templates are in absolute magnitude across frequencies.
  \[
  d_{\mathrm{rmse}}(x,y) = \sqrt{\frac{1}{F}\sum_f (x_f - y_f)^2}.
  \] 
\end{itemize}

\textbf{Objectives of prototype selection.}
Let $E=\{e_1,\ldots,e_N\}$ denote the set of environments represented by PSD templates. Our goal is to select a small subset of environments $P \subseteq E$ of size $k$, which will serve as prototypes. For any environment $e_i$, we measure its distance from the selected prototype set as
\[
d(e_i,P)=\min_{p\in P} d(e_i,p),
\]
which assigns each environment to its closest prototype.

Since the number of environments is small ($N=12$), we evaluate candidate prototype sets using exhaustive search. For each value of $k$, we enumerate all possible subsets $P \subseteq E$ with $|P|=k$ and evaluate each subset under both distance metrics. We consider two coverage objectives.

\begin{figure}[H]
\centering
\includegraphics[width=0.95\columnwidth]{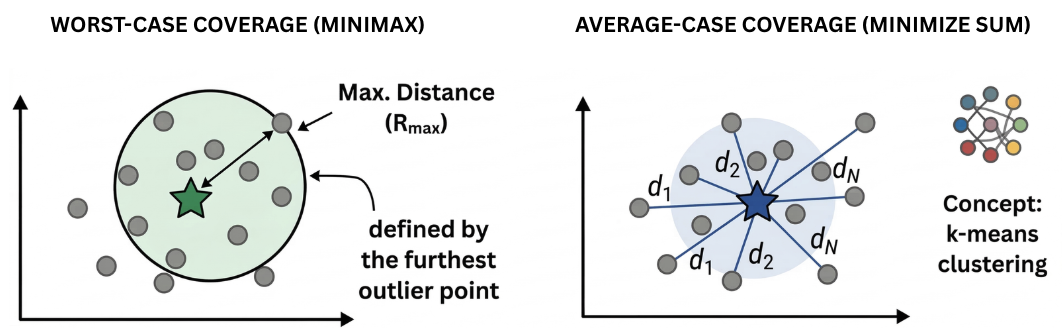}
\caption{Prototype-selection objectives: minimax (left) and average-case (right).}
\label{fig:proto_objectives}
\end{figure}

\emph{Worst-case coverage, or minimax coverage:}
As illustrated in the left panel of Fig.~\ref{fig:proto_objectives}, this objective is determined by the farthest environment from its nearest prototype. It encourages every environment to remain close to at least one selected prototype:
\[
\min_{P \subseteq E,\; |P|=k}
\max_{e_i \in E} d(e_i,P).
\]

\emph{Average-case coverage:}
As illustrated in the right panel of Fig.~\ref{fig:proto_objectives}, this objective depends on the mean distance from all environments to their nearest prototypes. It favors prototype sets that provide good overall coverage on average:
\begin{equation}
\min_{P \subseteq E,\; |P|=k}
\frac{1}{N}\sum_{e_i \in E} d(e_i,P).
\label{eq:average-coverage}
\end{equation}

This process is conceptually related to clustering methods such as $k$-means, where cluster centers represent groups of data points. In our setting, the selected environments act as prototype centers, and each environment is assigned to its nearest prototype according to the chosen PSD-template distance.

\begin{figure}[H]
\centering
\includegraphics[width=0.95\columnwidth]{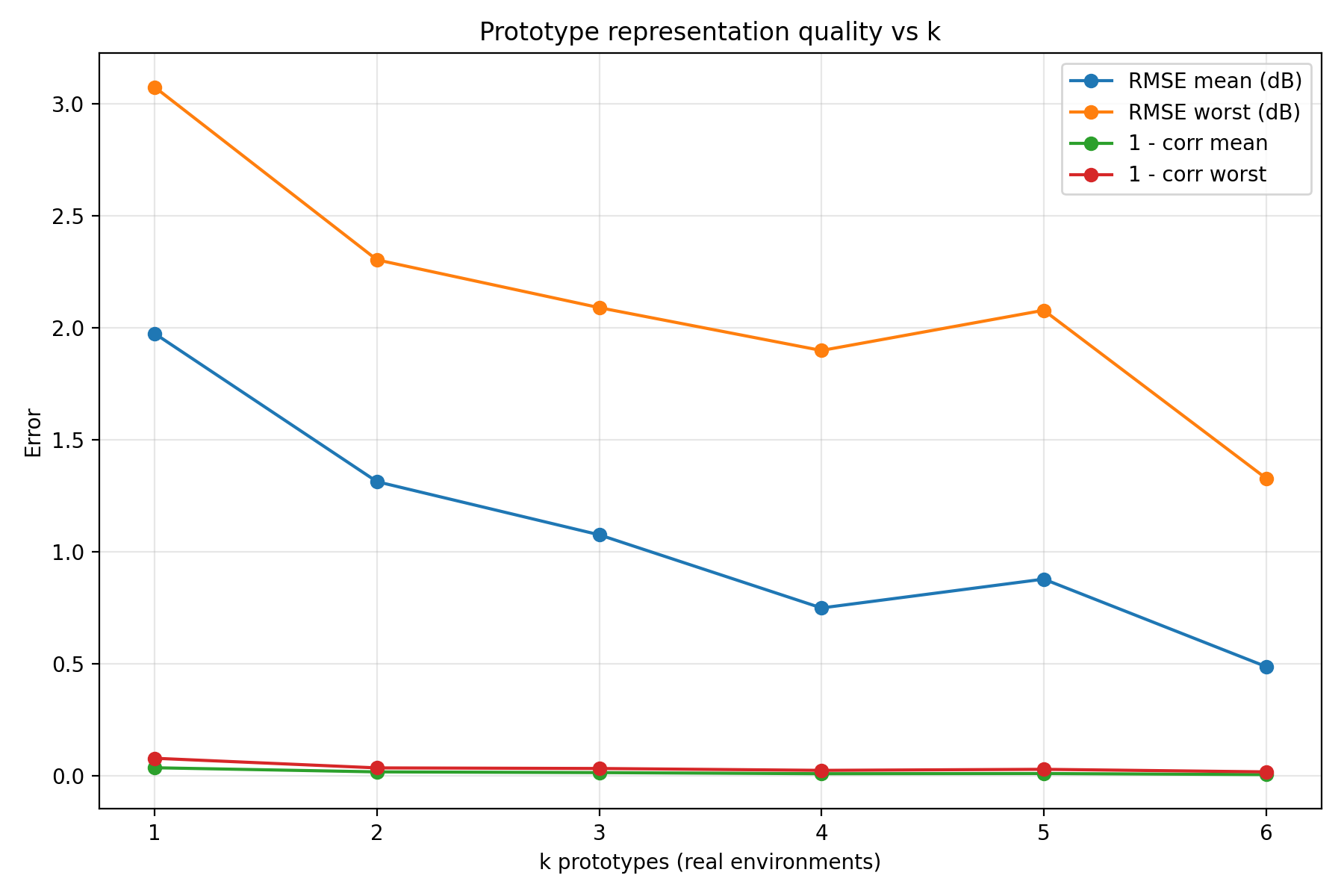}
\caption{Prototype representation quality vs.\ number of selected environments $k$. Both RMSE and correlation-distance exhibit diminishing returns after $k \in \{3,4\}$.}
\label{fig:proto-knee}
\end{figure}

\textbf{Selection of prototypes and choice of $k$.}
Using the coverage objectives defined above, we evaluate candidate prototype sets as a function of the number of selected environments $k$. As shown in Fig.~\ref{fig:proto-knee}, the largest reduction in representation error occurs between $k=1$ and $k=3$, after which improvements diminish. Based on worst-case coverage, we treat $k=3$ and $k=4$ as the main operating points. The set $k=3$ provides compact coverage suitable for efficient experimentation, while $k=4$ slightly improves worst-case coverage. The exact prototype sets and coverage metrics obtained through exhaustive evaluation are reported in supplementary material.

\section{Experimental Protocol}
\label{sec:exp}

We used the Indic-ST corpus \cite{sethiya2025indic}, focusing on 50 hours of English-Hindi speech for training and a 5-hour held-out evaluation set, with all audio processed at 16 kHz. We also use the same setting for English-German speech from the CoVoST dataset~\cite{wang2020covost}. We evaluate robustness on Whisper \cite{radford2022robustspeechrecognitionlargescale} and SeamlessM4T \cite{communication2023seamlessmultilingualexpressivestreaming}, fine-tuning medium-sized pretrained checkpoints and evaluating Hindi and German generation on clean, real-noise, and simulator-noise conditions. We report BLEU, chrF, and WER as evaluation metrics. Full training and decoding details are provided in supplementary material.

\section{Primary Robustness Comparison}
\label{subsec:main-compare}

We benchmark the primary robustness comparison on the English-Hindi dataset under five training noise conditions using the same 50-hour training budget. The regimes include clean-only training, Gaussian noise, pink noise, balanced real environment noise (uniformly sampled from 12 environments), and \simname-generated noise.

The key outcome is the close similarity between real-noise and simulator-noise training (Table~\ref{tab:main-robustness}). In SeamlessM4T, the simulator slightly improves WER relative to real-noise training while maintaining comparable BLEU and chrF. On Whisper, the two are essentially tied. These results show that \simname-generated noise achieves performance comparable to balanced real-noise training under the evaluated conditions.

\begin{table}[H]
\centering
\caption{Main robustness comparison (50h train, 5h test).}
\label{tab:main-robustness}
\resizebox{\columnwidth}{!}{%
\begin{tabular}{l
  S[table-format=2.3] S[table-format=1.4] S[table-format=2.3]
  S[table-format=2.3] S[table-format=1.4] S[table-format=2.3]}
\toprule
& \multicolumn{3}{c}{\textbf{Seamless}} & \multicolumn{3}{c}{\textbf{Whisper}} \\
\cmidrule(lr){2-4}\cmidrule(lr){5-7}
Training condition & {BLEU$\uparrow$} & {WER$\downarrow$} & {chrF$\uparrow$} &
{BLEU$\uparrow$} & {WER$\downarrow$} & {chrF$\uparrow$} \\
\midrule
Clean only      & 47.628 & 0.4322 & 66.400 & 34.168 & 0.5716 & 53.034 \\
Gaussian        & 46.523 & 0.4467 & 65.609 & 29.994 & 0.6265 & 46.340 \\
Pink (10 dB)    & 47.606 & 0.4310 & 66.848 & 34.019 & 0.5700 & 53.117 \\
Real env (12)   & 48.959 & 0.4293 & 67.893 & \bfseries 35.554 & 0.5564 & 54.143 \\
\simname\ noise & \bfseries 48.996 & \bfseries 0.4221 & \bfseries 67.896 & 35.494 & \bfseries 0.5549 & \bfseries 54.160 \\
\bottomrule
\end{tabular}}
\end{table}

As shown in Table~\ref{tab:baseline-gap}, Gaussian noise is a weak proxy for real environments, with especially large degradation on Whisper. Pink noise is stronger than Gaussian, but it still remains consistently worse than real-noise and simulator-noise training on both models. This suggests that matching only coarse spectral color is insufficient, and that the consistent gains of SIM noise over Gaussian and pink noise provide a proof of concept that more structured and realistic environment simulation is beneficial for robust training. It motivates simulator components that go beyond stationary PSD matching to envelope dynamics, transient activity, and chatter.

\begin{table}[H]
\centering
\caption{Performance gap relative to \simname\ noise. Negative BLEU/chrF and positive WER indicate worse performance than the simulator.}
\label{tab:baseline-gap}
\resizebox{\columnwidth}{!}{%
\begin{tabular}{l l r r r}
\toprule
Model & Baseline & {$\Delta$BLEU} & {$\Delta$WER} & {$\Delta$chrF} \\
\midrule
SeamlessM4T & Gaussian & \num{-2.473} & \num{0.0246} & \num{-2.287} \\
SeamlessM4T & Pink     & \num{-1.390} & \num{0.0089} & \num{-1.048} \\
Whisper     & Gaussian & \num{-5.500} & \num{0.0716} & \num{-7.820} \\
Whisper     & Pink     & \num{-1.475} & \num{0.0151} & \num{-1.043} \\
\bottomrule
\end{tabular}}
\end{table}

\begin{table}[H]
\centering
\caption{Effect of clean:noise mixture ratio. Entries are mean $\pm$ std over seeds.}
\label{tab:clean_noise_ratio_scaled}
\resizebox{\columnwidth}{!}{%
\begin{tabular}{l c c c}
\toprule
Setting & Noisy BLEU$\uparrow$ & Noisy WER$\downarrow$ & Noisy chrF$\uparrow$ \\
\midrule
clean20\_noise80 & \num{46.14}$\pm$\num{0.43} & \num{0.465}$\pm$\num{0.008} & \textbf{\num{66.63}}$\pm$\num{0.23} \\
clean40\_noise60 & \num{46.45}$\pm$\num{1.20} & \num{0.469}$\pm$\num{0.019} & \num{66.43}$\pm$\num{0.26} \\
clean50\_noise50 & \num{46.99}$\pm$\num{1.58} & \num{0.460}$\pm$\num{0.026} & \num{66.57}$\pm$\num{0.31} \\
clean70\_noise30 & \textbf{\num{47.12}}$\pm$\num{0.48} & \textbf{\num{0.457}}$\pm$\num{0.008} & \num{66.42}$\pm$\num{0.15} \\
\addlinespace[0.5ex] 
\midrule
Setting & Clean BLEU$\uparrow$ & Clean WER$\downarrow$ & Clean chrF$\uparrow$ \\
\midrule
clean20\_noise80 & \num{48.76}$\pm$\num{0.19} & \num{0.436}$\pm$\num{0.003} & \num{67.88}$\pm$\num{0.11} \\
clean40\_noise60 & \num{48.47}$\pm$\num{0.63} & \num{0.443}$\pm$\num{0.012} & \num{67.70}$\pm$\num{0.20} \\
clean50\_noise50 & \num{48.72}$\pm$\num{0.86} & \num{0.440}$\pm$\num{0.014} & \num{67.80}$\pm$\num{0.03} \\
clean70\_noise30 & \textbf{\num{49.91}}$\pm$\num{0.19} & \textbf{\num{0.425}}$\pm$\num{0.003} & \textbf{\num{67.95}}$\pm$\num{0.13} \\
\bottomrule
\end{tabular}}
\end{table}

\section{Impact of Simulator Components}
\label{sec:ablations}

We evaluate whether the simulator controls correspond to meaningful factors for robust training on SeamlessM4T. In each ablation, we keep the training pipeline fixed and vary only one control group.

\textbf{Chatter ON/OFF.}
Enabling chatter improves both noisy and clean performance (Table~\ref{tab:chatter-abl}), suggesting that speech-like interference is an important factor not well captured by stationary background noise alone.

\begin{table}[H]
\centering
\caption{Chatter ablation. Entries are mean $\pm$ std over seeds}
\label{tab:chatter-abl}
\resizebox{\columnwidth}{!}{%
\begin{tabular}{l c c c}
\toprule
Setting & Noisy BLEU$\uparrow$ & Noisy WER$\downarrow$ & Noisy chrF$\uparrow$ \\
\midrule
CHATTER\_OFF & \num{46.45}$\pm$\num{0.10} & \num{0.464}$\pm$\num{0.006} & \num{66.61}$\pm$\num{0.32} \\
CHATTER\_ON  & \textbf{\num{47.87}}$\pm$\textbf{\num{0.14}} & \textbf{\num{0.450}}$\pm$\textbf{\num{0.001}} & \textbf{\num{67.44}}$\pm$\textbf{\num{0.86}} \\
\addlinespace[0.5ex]
\midrule
Setting & Clean BLEU$\uparrow$ & Clean WER$\downarrow$ & Clean chrF$\uparrow$ \\
\midrule
CHATTER\_OFF & \num{48.24}$\pm$\num{0.24} & \num{0.445}$\pm$\num{0.002} & \num{67.63}$\pm$\num{0.40} \\
CHATTER\_ON  & \textbf{\num{49.36}}$\pm$\textbf{\num{0.03}} & \textbf{\num{0.431}}$\pm$\textbf{\num{0.000}} & \textbf{\num{68.52}}$\pm$\textbf{\num{0.17}} \\
\bottomrule
\end{tabular}}
\end{table}

\textbf{RIR ON/OFF.}
Enabling RIR reduces performance on both noisy and clean evaluation (Table~\ref{tab:all-ablations}), consistent with reverberation acting as a domain shift when training and test reverberation are mismatched \cite{7953152}. RIR should be treated as an explicit control rather than a default.
\begin{table}[H]
\centering
\renewcommand{\arraystretch}{0.60} 
\caption{Additional ablations across simulator components.}
\label{tab:all-ablations}
\resizebox{\columnwidth}{!}{%
\begin{tabular}{l c c c}
\toprule
Setting & Noisy BLEU$\uparrow$ & Noisy WER$\downarrow$ & Noisy chrF$\uparrow$ \\
\midrule
RIR off      & \textbf{45.86} & \textbf{0.4593} & \textbf{65.62} \\
RIR on       & 44.90 & 0.4798 & 64.36 \\
\midrule
Bursty       & 45.40 & 0.4676 & 65.10 \\
Continuous   & 45.48 & 0.4658 & 65.49 \\
\midrule
Full sim     & \textbf{46.97} & \textbf{0.4464} & \textbf{66.90} \\
PSD only     & 45.07 & 0.4645 & 65.78 \\
\addlinespace[0.5ex] 
\midrule
Setting & Clean BLEU$\uparrow$ & Clean WER$\downarrow$ & Clean chrF$\uparrow$ \\
\midrule
RIR off      & \textbf{47.73} & \textbf{0.4367} & \textbf{67.10} \\
RIR on       & 47.21 & 0.4542 & 65.96 \\
\midrule
Bursty       & 47.04 & 0.4482 & 66.80 \\
Continuous   & 47.53 & 0.4405 & 67.13 \\
\midrule
Full sim     & \textbf{48.94} & \textbf{0.4316} & \textbf{67.90} \\
PSD only     & 47.58 & 0.4399 & 67.10 \\
\bottomrule
\end{tabular}}
\renewcommand{\arraystretch}{1.0} 
\end{table}

\textbf{Continuous vs.\ bursty chatter.} The difference between continuous and bursty chatter is small, as shown in Table~\ref{tab:all-ablations}. This suggests that the presence of chatter and its overall level matter more than its fine-grained temporal scheduling in this setting.

\textbf{PSD-only vs.\ full simulator.} The PSD-only variant performs worse than the full simulator, shown in Table~\ref{tab:all-ablations}. This indicates that stationary spectral matching alone is insufficient; temporal dynamics and structured interference contribute meaningfully to robustness. 

\section{Hyperparameter Search Over Simulator Knobs}
\label{sec:hyp}

After establishing that the simulator does not degrade performance and testing the importance of the knobs, we conduct a structured sweep over the simulator hyperparameters to systematically assess how different knob configurations affect performance. We fine-tune only the SeamlessM4T model, as it performs better than Whisper based on previous results.

We conduct a structured sweep consisting of 27 experiments on both the language pair datasets: 9 simulator configurations (Table~\ref{tab:sweep-configs}) evaluated across 3 fitted PSD templates—cafe, washing (laundry), and car (Section~\ref{sec:proto}). Each configuration specifies a set of simulator controls governing background energy statistics, temporal dynamics, and speech-to-noise mixing. 

The simulator configurations are defined by a set of parameters; the full list and definitions are provided in supplementary material. We select the best configurations by minimizing the robustness objective
\begin{equation}
\mathrm{AvgWER}_{\mathrm{rob}} = \frac{1}{2} \left( \mathrm{WER}_{\mathrm{real}} + \mathrm{WER}_{\mathrm{sim}} \right),
\label{eq:avgwer-rob}
\end{equation}
while monitoring clean WER to avoid sacrificing clean accuracy.

\begin{table}[t]
\centering
\caption{Sweep configurations (shared across PSD templates).}
\label{tab:sweep-configs}
\resizebox{\columnwidth}{!}{%
\begin{tabular}{l
  S[table-format=-1.1]
  S[table-format=1.1]
  S[table-format=1.6]
  S[table-format=1.1]
  S[table-format=1.2]
  S[table-format=2.0]
  S[table-format=1.0]}
\toprule
Cfg & {$\mu_{\log \mathrm{RMS}}$} & {$\sigma_{\log \mathrm{RMS}}$} & {$r_e$ (Hz)} &
{$\lambda$} & {$\sigma_\lambda$} & {$\mu_{\mathrm{SNR}}$} & {$\sigma_{\mathrm{SNR}}$} \\
\midrule
cfg01 & -6.6 & 0.3 & 0.015259 & 2.2 & 0.30 & 0  & 3 \\
cfg02 & -6.6 & 0.3 & 0.045776 & 3.6 & 0.60 & 10 & 4 \\
cfg03 & -6.6 & 0.3 & 0.106812 & 4.8 & 0.25 & 20 & 5 \\
cfg04 & -5.2 & 0.5 & 0.015259 & 3.6 & 0.60 & 20 & 5 \\
cfg05 & -5.2 & 0.5 & 0.045776 & 4.8 & 0.25 & 0  & 3 \\
cfg06 & -5.2 & 0.5 & 0.106812 & 2.2 & 0.30 & 10 & 4 \\
cfg07 & -3.3 & 0.7 & 0.015259 & 4.8 & 0.25 & 10 & 4 \\
cfg08 & -3.3 & 0.7 & 0.045776 & 2.2 & 0.30 & 20 & 5 \\
cfg09 & -3.3 & 0.7 & 0.106812 & 3.6 & 0.60 & 0  & 3 \\
\bottomrule
\end{tabular}}
\end{table}

\subsection{Sweep results and selected default}

We summarize the sweep outcomes by aggregating performance across the three PSD templates. The goal is to identify configurations that perform consistently well under the robustness objective in Equation~\ref{eq:avgwer-rob}.

As shown in Table~\ref{tab:sweep-main} and Table~\ref{tab:sweep-main-ende}, configuration 3 (cfg03) and configuration 6 (cfg06) achieve the strongest robustness outcomes in this sweep, with consistently low variance across environments:

\begin{itemize}[leftmargin=*, itemsep=0pt, parsep=0pt, topsep=2pt]
    \item \textbf{cfg03:} $\mu_{\log \mathrm{RMS}} = -6.6$, $\sigma_{\log \mathrm{RMS}} = 0.3$, $r_e = 0.106812\,\mathrm{Hz}$, $\lambda = 4.8$, $\sigma_\lambda = 0.25$, SNR $\sim \mathcal{N}(20,5)$ dB.
    \item \textbf{cfg06:} $\mu_{\log \mathrm{RMS}} = -5.2$, $\sigma_{\log \mathrm{RMS}} = 0.5$, $r_e = 0.106812\,\mathrm{Hz}$, $\lambda = 2.2$, $\sigma_\lambda = 0.30$, SNR $\sim \mathcal{N}(10,4)$ dB.
\end{itemize}

\begin{table}[H]
\centering
\caption{Sweep results aggregated over PSD templates (cafe, washing, car). Entries are mean $\pm$ std over environments.}
\label{tab:sweep-main}
\resizebox{\columnwidth}{!}{%
\begin{tabular}{l c c c c}
\toprule
Cfg & {Clean WER$\downarrow$} & {Real WER$\downarrow$} & {Sim WER$\downarrow$} & {$\mathrm{AvgWER}_{\mathrm{rob}}\downarrow$} \\
\midrule
cfg01 & \num{0.4271}$\pm$\num{0.0026} & \num{0.4500}$\pm$\num{0.0052} & \num{0.4424}$\pm$\num{0.0020} & \num{0.4462}$\pm$\num{0.0031} \\
cfg02 & \num{0.4250}$\pm$\num{0.0016} & \num{0.4435}$\pm$\num{0.0046} & \num{0.4426}$\pm$\num{0.0106} & \num{0.4430}$\pm$\num{0.0073} \\
\textbf{cfg03} &
\textbf{\num{0.4217}}$\pm$\textbf{\num{0.0013}} &
\textbf{\num{0.4371}}$\pm$\textbf{\num{0.0049}} &
\textbf{\num{0.4386}}$\pm$\textbf{\num{0.0095}} &
\textbf{\num{0.4379}}$\pm$\textbf{\num{0.0072}} \\
cfg04 & \num{0.4344}$\pm$\num{0.0115} & \num{0.4517}$\pm$\num{0.0162} & \num{0.4458}$\pm$\num{0.0161} & \num{0.4487}$\pm$\num{0.0161} \\
cfg05 & \num{0.4323}$\pm$\num{0.0090} & \num{0.4488}$\pm$\num{0.0087} & \num{0.4472}$\pm$\num{0.0070} & \num{0.4480}$\pm$\num{0.0076} \\
\textbf{cfg06} &
\textbf{\num{0.4278}}$\pm$\textbf{\num{0.0012}} &
\textbf{\num{0.4365}}$\pm$\textbf{\num{0.0010}} &
\textbf{\num{0.4333}}$\pm$\textbf{\num{0.0025}} &
\textbf{\num{0.4349}}$\pm$\textbf{\num{0.0009}} \\
cfg07 & \num{0.4380}$\pm$\num{0.0075} & \num{0.4556}$\pm$\num{0.0054} & \num{0.4561}$\pm$\num{0.0130} & \num{0.4559}$\pm$\num{0.0087} \\
cfg08 & \num{0.4260}$\pm$\num{0.0128} & \num{0.4499}$\pm$\num{0.0089} & \num{0.4428}$\pm$\num{0.0098} & \num{0.4464}$\pm$\num{0.0084} \\
cfg09 & \num{0.4380}$\pm$\num{0.0098} & \num{0.4541}$\pm$\num{0.0092} & \num{0.4490}$\pm$\num{0.0082} & \num{0.4516}$\pm$\num{0.0085} \\
\bottomrule
\end{tabular}}
\end{table}

\begin{table}[H]
\centering
\caption{Sweep results for English to German aggregated over PSD templates. Entries are mean $\pm$ std over environments.}
\label{tab:sweep-main-ende}
\resizebox{\columnwidth}{!}{%
\begin{tabular}{l c c c c}
\toprule
Cfg & {Clean WER$\downarrow$} & {Real WER$\downarrow$} & {Sim WER$\downarrow$} & {$\mathrm{AvgWER}_{\mathrm{rob}}\downarrow$} \\
\midrule
cfg01 & \num{0.554}$\pm$\num{0.006} & \num{0.580}$\pm$\num{0.003} & \num{0.582}$\pm$\num{0.004} & \num{0.581}$\pm$\num{0.003} \\
cfg02 & \num{0.553}$\pm$\num{0.003} & \num{0.582}$\pm$\num{0.003} & \num{0.583}$\pm$\num{0.002} & \num{0.583}$\pm$\num{0.003} \\
\textbf{cfg03} & \textbf{\num{0.549}}$\pm$\textbf{\num{0.002}} & \textbf{\num{0.578}}$\pm$\textbf{\num{0.001}} & \textbf{\num{0.580}}$\pm$\textbf{\num{0.003}} & \textbf{\num{0.579}}$\pm$\textbf{\num{0.002}} \\
cfg04 & \num{0.553}$\pm$\num{0.006} & \num{0.581}$\pm$\num{0.006} & \num{0.583}$\pm$\num{0.004} & \num{0.582}$\pm$\num{0.005} \\
cfg05 & \num{0.552}$\pm$\num{0.002} & \num{0.581}$\pm$\num{0.004} & \num{0.582}$\pm$\num{0.004} & \num{0.581}$\pm$\num{0.004} \\
\textbf{cfg06} & \textbf{\num{0.550}}$\pm$\textbf{\num{0.001}} & \textbf{\num{0.579}}$\pm$\textbf{\num{0.002}} & \textbf{\num{0.581}}$\pm$\textbf{\num{0.000}} & \textbf{\num{0.580}}$\pm$\textbf{\num{0.001}} \\
cfg07 & \num{0.552}$\pm$\num{0.004} & \num{0.579}$\pm$\num{0.003} & \num{0.582}$\pm$\num{0.005} & \num{0.581}$\pm$\num{0.004} \\
cfg08 & \num{0.552}$\pm$\num{0.006} & \num{0.581}$\pm$\num{0.005} & \num{0.585}$\pm$\num{0.008} & \num{0.583}$\pm$\num{0.007} \\
cfg09 & \num{0.551}$\pm$\num{0.002} & \num{0.580}$\pm$\num{0.005} & \num{0.581}$\pm$\num{0.001} & \num{0.580}$\pm$\num{0.003} \\
\bottomrule
\end{tabular}}
\end{table}

\begin{figure}[H]
\centering
\includegraphics[width=0.90\columnwidth]{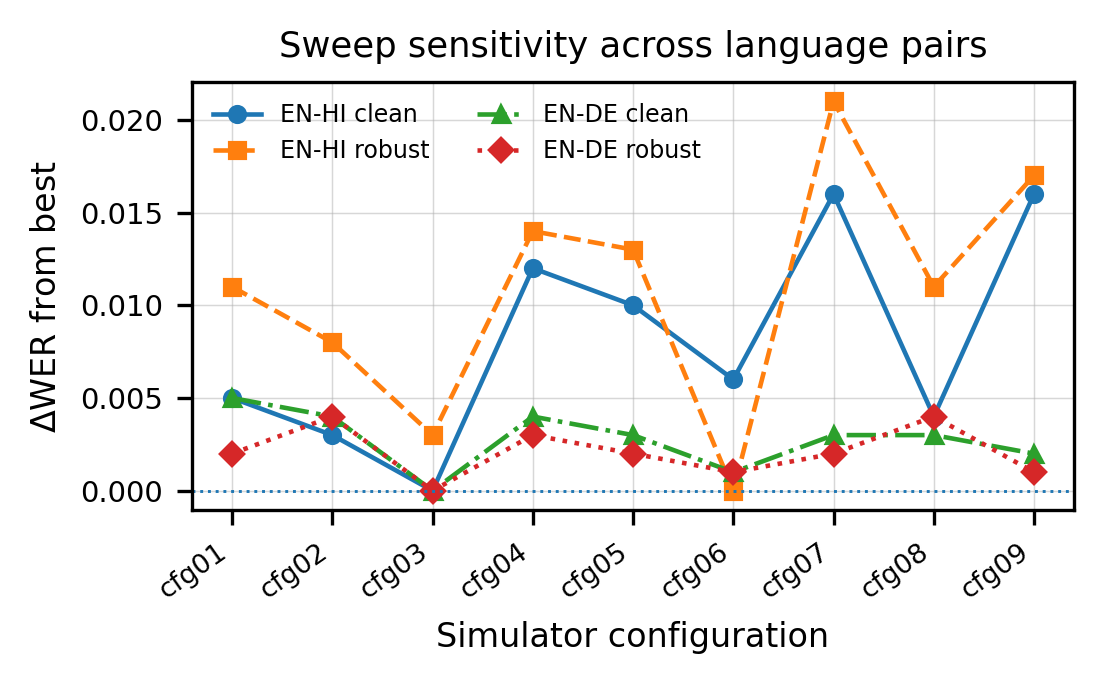}
\caption{Sweep sensitivity across language pairs. Values show the WER increase relative to the best configuration for each metric and language pair, using the clean WER and $\mathrm{AvgWER}_{\mathrm{rob}}$ columns from Tables~\ref{tab:sweep-main} and~\ref{tab:sweep-main-ende}. Lower is better.}
\label{result_plot_compact}
\end{figure}

Fig.~\ref{result_plot_compact} shows that cfg03 and cfg06 are the most stable choices across both language pairs and across clean and robust evaluation. We do not claim that cfg03 or cfg06 is globally optimal. Rather, they are strong defaults within this small structured search space and across the three PSD templates studied. The comparison suggests that faster envelope dynamics ($r_e$) is consistently helpful, while robustness can be obtained either through a higher-SNR level with higher event rates (cfg03) or through a moderate-SNR level with lower event rates (cfg06). These trends are empirical for our setting and may change across deployment conditions.

To examine whether the selected default varies by environment, we also report the best run for each PSD template separately. 

\begin{table}[t]
\centering
\caption{Best run per PSD template (min $\mathrm{AvgWER}_{\mathrm{rob}}$) for en-hi.}
\label{tab:best-per-env_hi}
\resizebox{\columnwidth}{!}{%
\begin{tabular}{l l
  S[table-format=1.4]
  S[table-format=1.4]
  S[table-format=1.4]
  S[table-format=1.4]}
\toprule
Env & Best exp & {Clean WER} & {Real WER} & {Sim WER} & {$\mathrm{AvgWER}_{\mathrm{rob}}$} \\
\midrule
cafe    & exp02 & 0.4238 & 0.4382 & 0.4334 & 0.4358 \\
washing & exp12 & 0.4204 & 0.4317 & 0.4285 & 0.4301 \\
car     & exp24 & 0.4264 & 0.4375 & 0.4305 & 0.4340 \\
\bottomrule
\end{tabular}}
\end{table}

\begin{table}[t]
\centering
\caption{Best run per PSD template (min $\mathrm{AvgWER}_{\mathrm{rob}}$) for en-de.}
\label{tab:best-per-env_de}
\resizebox{\columnwidth}{!}{%
\begin{tabular}{l l
  S[table-format=1.6]
  S[table-format=1.6]
  S[table-format=1.6]
  S[table-format=1.6]}
\toprule
Env & Best exp & {Clean WER} & {Real WER} & {Sim WER} & {$\mathrm{AvgWER}_{\mathrm{rob}}$} \\
\midrule
cafe    & exp04 & 0.554214 & 0.578458 & 0.580612 & 0.579535 \\
washing & exp12 & 0.547752 & 0.575962 & 0.580031 & 0.577997 \\
car     & exp19 & 0.547205 & 0.576611 & 0.576817 & 0.576714 \\
\bottomrule
\end{tabular}}
\end{table}

The best configuration varies slightly by PSD template, as shown in Table~\ref{tab:best-per-env_hi} and Table~\ref{tab:best-per-env_de}, which is expected because different spectral shapes and temporal properties imply different masking regimes. In practice, cfg03 and cfg06 provide good general defaults, while per-environment tuning can be treated as a second-stage refinement when deployment is dominated by a single environment type.
\section{Discussion}
\label{sec:discussion}

The central design choice in \framework\ is \emph{controllability}, so that changes in robustness can be attributed to explicit simulator controls or selection objectives. Training with simulator-generated noise matches balanced real-noise training under the same data budget and evaluation conditions, indicating that the simulator captures the statistics that matter for robustness in the studied settings. This framework highlights that prototype selection is most useful when the environment catalog is large and training budgets are limited. Compute PSD templates, select $k$ environments via the coverage objective, and focus fitting and tuning on those prototypes. A conservative workflow is to start with $k=3$ prototypes for rapid iteration, fit simulator targets from real recordings for those prototypes, train with cfg03 or cfg06 as robust default configurations, and then use ablations to decide whether chatter and RIR are appropriate for the intended deployment domain. When deployment is dominated by a single environment type, the same procedure can be refined by tuning knobs per PSD template using the per-environment selection approach.

\section{Conclusion}
\label{sec:conclusion}

We introduced \framework, a structured robustness workflow for speech-to-text training that combines a controllable environment simulator, coverage-optimal environment reduction on PSD templates, and a structured hyperparameter search over simulator knobs. We demonstrated performance comparable to balanced real-noise training across Whisper and SeamlessM4T. Through ablations, we showed that dynamics matter beyond stationary spectral matching and provided principled environment prototype sets. Finally, we selected practical default simulator configurations from a 27-run sweep. Together, these components form a reproducible and interpretable path toward robust S2TT training.

\newpage



\section*{Acknowledgment}

\textbf{AI-Generated Content Disclosure:} The authors used AI-based writing and research-assistance tools, including OpenAI ChatGPT, Anthropic Claude Sonnet, Google Gemini 3.1 Pro, and xAI Grok 4.3, to assist with manuscript review, language polishing, grammar correction, organization suggestions, LaTeX formatting suggestions, and drafting of explanatory text. AI assistance was applied to portions of the Abstract, Introduction, Related Work, Method description, Experimental Protocol, Discussion, Conclusion, and Acknowledgment sections.

The AI systems were not used to fabricate experimental results, datasets, evaluation metrics, figures, tables, or citations. All experimental design choices, simulator implementation, training runs, evaluation results, analyses, and final scientific claims were reviewed, edited, and verified by the authors, who take full responsibility for the final content of the paper.


\clearpage
\onecolumn

\section*{Supplementary Material}

\appendices

\section{Method Details}
\label{supp:method_details}

\subsection{Derivation of Noise Scaling Factor}
\label{supp:snr_derivation}

Let $s(t)$ denote the clean speech signal and $x_e(t)$ the simulated
ambience. The mixed signal is
\begin{equation}
y(t) = s(t) + g\,x_e(t),
\end{equation}
where $g$ is chosen to achieve a target signal-to-noise ratio (SNR).

We define the SNR in decibels using RMS amplitudes:
\begin{equation}
\mathrm{SNR}
=
20\log_{10}
\left(
\frac{\RMS(s)}{\RMS(gx_e)}
\right).
\end{equation}

Using the scaling property of RMS,
\begin{equation}
\RMS(gx_e) = g\,\RMS(x_e),
\end{equation}
we obtain
\begin{align}
\mathrm{SNR}
&=
20\log_{10}
\left(
\frac{\RMS(s)}{g\,\RMS(x_e)}
\right), \\
\frac{\mathrm{SNR}}{20}
&=
\log_{10}
\left(
\frac{\RMS(s)}{g\,\RMS(x_e)}
\right), \\
10^{\mathrm{SNR}/20}
&=
\frac{\RMS(s)}{g\,\RMS(x_e)}.
\end{align}

Solving for $g$ gives
\begin{equation}
g =
\frac{\RMS(s)}{\RMS(x_e)}
10^{-\mathrm{SNR}/20}.
\end{equation}

This is the scaling factor used during speech-noise mixing in
\simname.

\section{Experimental Details}
\label{supp:experimental_details}

\subsection{Training and Decoding Details}
\label{supp:training_details}

Table~\ref{tab:supp_training_details} summarizes the principal
training and decoding settings used for the evaluated models.

\begin{table}[H]
\centering
\caption{Training and decoding settings for SeamlessM4T and Whisper.}
\label{tab:supp_training_details}
\setlength{\tabcolsep}{8pt}
\begin{tabular}{p{0.28\textwidth} p{0.31\textwidth} p{0.31\textwidth}}
\toprule
\textbf{Setting} & \textbf{SeamlessM4T} & \textbf{Whisper} \\
\midrule
Checkpoint
& hf-seamless-m4t-medium
& whisper-medium \\

Trainer
& HuggingFace Trainer
& HuggingFace Seq2SeqTrainer \\

Audio sampling rate
& 16 kHz
& 16 kHz \\

Batch size
& 4
& 8 \\

Gradient accumulation
& 4
& 4 \\

Optimizer
& AdamW
& AdamW \\

Learning rate
& $3\times10^{-5}$
& $3\times10^{-5}$ \\

Scheduler
& Cosine
& Cosine \\

Warmup ratio
& 0.1
& 0.1 \\

Weight decay
& 0.03
& 0.03 \\

Maximum gradient norm
& 0.5
& 0.5 \\

Epoch budget
& Up to 10
& Up to 10 \\

Early stopping
& Patience 5, threshold 0.005
& Patience 5, threshold 0.005 \\

Precision
& fp16
& bf16 if available, otherwise fp16 \\

Gradient checkpointing
& Yes
& Yes \\

Target decoding language
& Hindi (en-hi); German (en-de)
& Hindi (en-hi); German (en-de) \\

Beam size
& 4
& 4 \\

Maximum new tokens
& 160
& 160 \\
\bottomrule
\end{tabular}
\end{table}

\subsection{Coverage-Based Environment Reduction}
\label{supp:coverage_reduction}

Table~\ref{tab:supp_proto} reports the environment prototypes obtained
under the PSD-based coverage analysis described in the main paper.
The prototypes are actual environments rather than synthetic cluster
centroids.

\begin{table}[H]
\centering
\caption{Prototype environments selected using PSD coverage.}
\label{tab:supp_proto}
\setlength{\tabcolsep}{7pt}
\begin{tabular}{c l c c c c c c}
\toprule
$k$
& Selected prototypes
& max $\distcorr$
& mean $\distcorr$
& RMSE$_\mu$ (dB)
& RMSE$_{\max}$ (dB)
& Corr$_\mu$
& Corr$_{\min}$ \\
\midrule
1
& bus
& 0.077 & 0.034 & 1.986 & 3.073 & 0.966 & 0.923 \\

2
& cafe, office
& 0.040 & 0.021 & 1.316 & 2.302 & 0.979 & 0.966 \\

3
& cafe, washing, car
& 0.032 & 0.015 & 1.074 & 2.089 & 0.987 & 0.969 \\

4
& washing, station, car, meeting
& 0.028 & 0.012 & 0.748 & 1.897 & 0.992 & 0.977 \\

5
& cafe, living, washing, bus, meeting
& 0.027 & 0.010 & 0.670 & 2.077 & 0.994 & 0.973 \\

6
& living, washing, field, station, car, meeting
& 0.020 & 0.008 & 0.489 & 1.327 & 0.996 & 0.984 \\
\bottomrule
\end{tabular}
\end{table}

Increasing $k$ improves spectral coverage of the environment set.
The reduction is largest over the first few prototypes and becomes
smaller after approximately $k=3$--$4$, consistent with the knee
observed in the coverage curve in the main paper. We therefore use
the $k=3$ set, consisting of cafe, washing, and car, as the compact
set for the subsequent simulator sweep.

\subsection{Full Simulator Parameter Sweep}
\label{supp:full_parameter_sweep}

For each language pair, the structured sweep contains nine simulator
configurations evaluated over three PSD templates: cafe, washing, and
car. This gives 27 experiments per language pair.

The full English-to-Hindi run-level results are reported below.
To maintain readable text, the results are divided into clean/real-noise
and simulator-noise tables rather than compressed into a single
extremely wide table.

\begin{table}[H]
\centering
\caption{Full English-to-Hindi sweep: clean and real-noise evaluation.}
\label{tab:supp_full_sweep_clean_real}
\setlength{\tabcolsep}{5pt}
\begin{tabular}{lllcccccc}
\toprule
Env & Exp & Cfg &
Clean BLEU & Clean chrF & Clean WER &
Real BLEU & Real chrF & Real WER \\
\midrule

cafe & exp01 & cfg01 & 48.668 & 67.761 & 0.425 & 47.421 & 66.771 & 0.444 \\
cafe & exp02 & cfg02 & 48.762 & 68.015 & 0.424 & 47.736 & 66.782 & 0.438 \\
cafe & exp03 & cfg03 & 48.959 & 67.903 & 0.423 & 47.422 & 66.582 & 0.438 \\
cafe & exp04 & cfg04 & 47.807 & 68.204 & 0.447 & 46.102 & 66.876 & 0.470 \\
cafe & exp05 & cfg05 & 47.869 & 67.975 & 0.443 & 46.771 & 66.912 & 0.457 \\
cafe & exp06 & cfg06 & 48.696 & 67.701 & 0.429 & 47.923 & 66.819 & 0.436 \\
cafe & exp07 & cfg07 & 48.491 & 67.617 & 0.435 & 47.037 & 66.399 & 0.456 \\
cafe & exp08 & cfg08 & 49.695 & 68.499 & 0.412 & 47.612 & 67.057 & 0.440 \\
cafe & exp09 & cfg09 & 47.261 & 67.436 & 0.449 & 46.143 & 66.484 & 0.463 \\

\midrule

washing & exp10 & cfg01 & 48.315 & 68.126 & 0.430 & 46.949 & 67.160 & 0.452 \\
washing & exp11 & cfg02 & 49.264 & 68.164 & 0.424 & 47.855 & 66.812 & 0.446 \\
washing & exp12 & cfg03 & 49.125 & 68.263 & 0.420 & 47.745 & 66.681 & 0.432 \\
washing & exp13 & cfg04 & 48.520 & 67.527 & 0.433 & 47.059 & 66.057 & 0.443 \\
washing & exp14 & cfg05 & 48.445 & 68.079 & 0.428 & 47.361 & 67.011 & 0.440 \\
washing & exp15 & cfg06 & 48.642 & 68.168 & 0.429 & 47.983 & 67.127 & 0.436 \\
washing & exp16 & cfg07 & 47.692 & 67.853 & 0.446 & 46.554 & 66.637 & 0.461 \\
washing & exp17 & cfg08 & 48.764 & 67.964 & 0.428 & 46.976 & 66.352 & 0.454 \\
washing & exp18 & cfg09 & 48.325 & 67.860 & 0.429 & 47.144 & 66.718 & 0.445 \\

\midrule

car & exp19 & cfg01 & 48.480 & 68.323 & 0.426 & 46.801 & 66.980 & 0.454 \\
car & exp20 & cfg02 & 48.777 & 67.943 & 0.427 & 47.328 & 66.603 & 0.447 \\
car & exp21 & cfg03 & 49.161 & 67.941 & 0.422 & 47.349 & 66.567 & 0.441 \\
car & exp22 & cfg04 & 48.911 & 68.028 & 0.424 & 47.445 & 66.541 & 0.442 \\
car & exp23 & cfg05 & 48.638 & 67.574 & 0.427 & 47.131 & 66.293 & 0.450 \\
car & exp24 & cfg06 & 48.852 & 68.243 & 0.426 & 47.752 & 66.862 & 0.438 \\
car & exp25 & cfg07 & 48.664 & 68.078 & 0.432 & 47.051 & 66.564 & 0.450 \\
car & exp26 & cfg08 & 48.169 & 67.857 & 0.438 & 46.718 & 66.479 & 0.456 \\
car & exp27 & cfg09 & 48.223 & 68.011 & 0.436 & 46.984 & 66.863 & 0.455 \\

\bottomrule
\end{tabular}
\end{table}

\begin{table}[H]
\centering
\caption{Full English-to-Hindi sweep: simulator-noise evaluation.}
\label{tab:supp_full_sweep_sim}
\setlength{\tabcolsep}{9pt}
\begin{tabular}{lllccc}
\toprule
Env & Exp & Cfg &
Sim BLEU & Sim chrF & Sim WER \\
\midrule

cafe & exp01 & cfg01 & 47.707 & 66.971 & 0.441 \\
cafe & exp02 & cfg02 & 48.112 & 67.316 & 0.433 \\
cafe & exp03 & cfg03 & 47.487 & 66.710 & 0.440 \\
cafe & exp04 & cfg04 & 46.550 & 67.108 & 0.464 \\
cafe & exp05 & cfg05 & 46.913 & 67.113 & 0.455 \\
cafe & exp06 & cfg06 & 47.955 & 66.902 & 0.435 \\
cafe & exp07 & cfg07 & 46.512 & 66.485 & 0.466 \\
cafe & exp08 & cfg08 & 47.726 & 67.340 & 0.437 \\
cafe & exp09 & cfg09 & 46.450 & 66.725 & 0.458 \\

\midrule

washing & exp10 & cfg01 & 47.294 & 67.329 & 0.445 \\
washing & exp11 & cfg02 & 47.914 & 66.940 & 0.440 \\
washing & exp12 & cfg03 & 48.077 & 67.086 & 0.429 \\
washing & exp13 & cfg04 & 47.414 & 66.524 & 0.438 \\
washing & exp14 & cfg05 & 47.256 & 67.191 & 0.442 \\
washing & exp15 & cfg06 & 48.097 & 67.359 & 0.434 \\
washing & exp16 & cfg07 & 46.540 & 66.982 & 0.461 \\
washing & exp17 & cfg08 & 47.751 & 66.959 & 0.437 \\
washing & exp18 & cfg09 & 47.475 & 67.104 & 0.443 \\

\midrule

car & exp19 & cfg01 & 47.416 & 67.355 & 0.441 \\
car & exp20 & cfg02 & 46.788 & 66.647 & 0.454 \\
car & exp21 & cfg03 & 47.350 & 66.523 & 0.447 \\
car & exp22 & cfg04 & 47.774 & 66.792 & 0.435 \\
car & exp23 & cfg05 & 47.520 & 66.540 & 0.444 \\
car & exp24 & cfg06 & 48.148 & 67.303 & 0.430 \\
car & exp25 & cfg07 & 47.563 & 66.882 & 0.441 \\
car & exp26 & cfg08 & 46.970 & 66.703 & 0.454 \\
car & exp27 & cfg09 & 47.647 & 67.278 & 0.445 \\

\bottomrule
\end{tabular}
\end{table}

Each experiment combines one of the nine configurations with one
of the three PSD templates. These run-level results are aggregated
by configuration in the main paper to compute clean, real-noise,
simulator-noise, and robustness-oriented WER statistics.

\section{Data and Environment Details}
\label{supp:data_environment_details}

\subsection{DEMAND Environment Categories}
\label{supp:demand_envs}

The DEMAND dataset contains recordings from 18 real-world acoustic
environments spanning indoor, outdoor, public, and transportation
conditions. The environment categories considered in our data
preparation are:

\begin{itemize}[leftmargin=*]
    \item \textbf{Indoor:} kitchen, living room, washing area, hallway,
    meeting room, office, cafeteria, and restaurant.
    \item \textbf{Outdoor and public spaces:} open field, park, river,
    station, street cafe, and public square.
    \item \textbf{Transportation and traffic:} traffic, bus, car, and metro.
\end{itemize}

The main experiments use a fixed subset of 12 environments from this
catalog for the real-noise comparison and PSD-based environment
reduction. The prototype-selection procedure is therefore performed
with $N=12$, as stated in the main paper. The subsequent structured
hyperparameter sweep uses the three selected PSD templates:
cafe, washing, and car.


\subsection{Spectral Examples from DEMAND}
\label{supp:psd_examples}

\begin{figure}[H]
\centering
\includegraphics[width=0.85\textwidth]{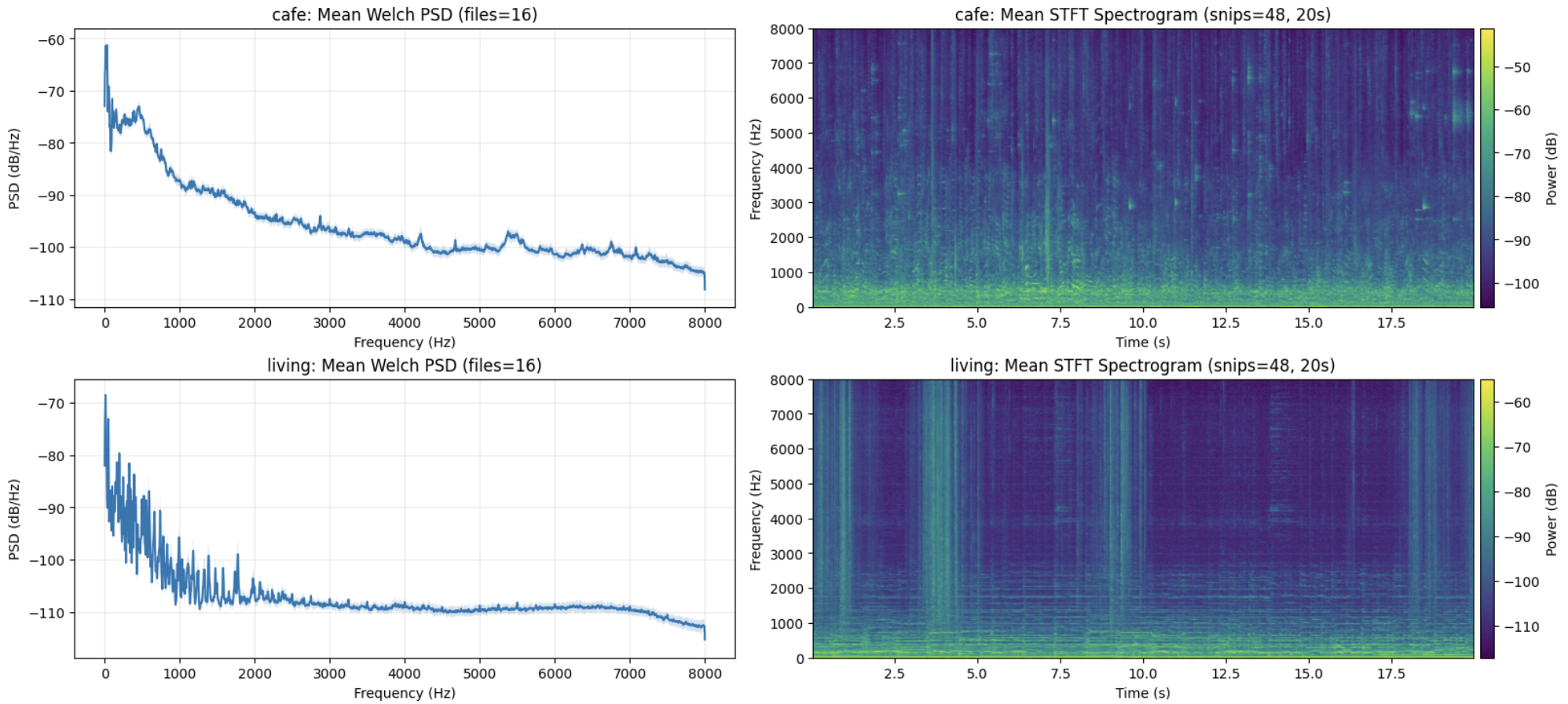}
\caption{Qualitative examples of acoustic structure for two DEMAND
environments, shown using mean Welch PSD curves and mean STFT
spectrograms.}
\label{fig:supp_psd_stft_examples}
\end{figure}

Figure~\ref{fig:supp_psd_stft_examples} provides qualitative examples
of acoustic structure in two DEMAND environments using mean Welch
PSD curves and mean STFT spectrograms. The PSD summarizes the
long-term frequency-wise energy distribution, while the spectrogram
shows how spectral energy evolves over time. The examples illustrate
that different environments exhibit distinct spectral profiles and
temporal structures, motivating the use of environment-specific PSD
templates in \framework.

\subsection{Sweep Parameter Definitions}
\label{supp:sweep_params}

The simulator configurations are defined using the following
parameters:

\begin{itemize}[leftmargin=*, itemsep=2pt]

    \item $\mu_{\log \mathrm{RMS}}$:
    mean log-RMS level controlling the characteristic background
    noise energy.

    \item $\sigma_{\log \mathrm{RMS}}$:
    variability of the log-RMS background energy.

    \item $r_e$:
    rate parameter controlling slow temporal variation of the
    ambience envelope.

    \item $\lambda$:
    mean rate of transient acoustic events.

    \item $\sigma_\lambda$:
    variability of the transient-event rate.

    \item $\mu_{\mathrm{SNR}}$:
    mean SNR used when mixing clean speech with environmental noise.

    \item $\sigma_{\mathrm{SNR}}$:
    variability of the sampled SNR across utterances.

\end{itemize}

\end{document}